\documentclass[conference]{IEEEtran}
\IEEEoverridecommandlockouts
\usepackage{cite}
\usepackage{amsmath,amssymb,amsfonts}
\usepackage{algorithmic}
\usepackage{graphicx}
\usepackage{textcomp}
\usepackage{booktabs}
\usepackage{multirow}
\usepackage[dvipsnames]{xcolor}
\usepackage{comment}
\usepackage{subcaption}
\usepackage{mwe} 

\def\BibTeX{{\rm B\kern-.05em{\sc i\kern-.025em b}\kern-.08em
    T\kern-.1667em\lower.7ex\hbox{E}\kern-.125emX}}

\usepackage{personalcommand}

\newcommand{\expComp}[1]{\textbf{#1} UML Components\xspace}
\newcommand{\expNode}[1]{\textbf{#1} UML Nodes\xspace}
\newcommand{\expUC}[1]{\textbf{#1} UML Use Cases\xspace}
\newcommand{\referenceP}{$PF^{ref}$\xspace}
\newcommand{\computedP}{$PF^{c}$\xspace}
\newcommand{\independentRun}{31\xspace}

\begin{document}

\title{Search Budget in Multi-Objective Refactoring Optimization: a Model-Based Empirical Study}

\author{\IEEEauthorblockN{Daniele Di Pompeo}
\IEEEauthorblockA{\textit{DISIM} \\
\textit{University of L'Aquila}\\
L'Aquila, Italy \\
daniele.dipompeo@univaq.it}
\and
\IEEEauthorblockN{Michele Tucci}
\IEEEauthorblockA{\textit{D3S} \\
\textit{Charles University}\\
Prague, Czech Republic \\
tucci@d3s.mff.cuni.cz}
}

\maketitle

\begin{abstract}
Software model optimization is the task of automatically generate design alternatives, usually to improve quality aspects of software that are quantifiable, like performance and reliability.
In this context, multi-objective optimization techniques have been applied to help the designer find suitable trade-offs among several non-functional properties.
In this process, design alternatives can be generated through automated model refactoring, and evaluated on non-functional models.
Due to their complexity, this type of optimization tasks require considerable time and resources, often limiting their application in software engineering processes. 

In this paper, we investigate the effects of using a search budget, specifically a time limit, to the search for new solutions.
We performed experiments to quantify the impact that a change in the search budget may have on the quality of solutions.
Furthermore, we analyzed how different genetic algorithms (\ie \nsga, \spea, and \pesa) perform when imposing different budgets.
We experimented on two case studies of different size, complexity, and domain.

We observed that imposing a search budget considerably deteriorates the quality of the generated solutions, but the specific algorithm we choose seems to play a crucial role.
From our experiments, \nsga is the fastest algorithm, while \pesa generates solutions with the highest quality.
Differently, \spea is the slowest algorithm, and produces the solutions with the lowest quality.
  
\end{abstract}


\section{Introduction}\label{sec:intro}

In the last decades, multi-objective optimization techniques have been successfully applied to many model-driven software development problems~\cite{Ramirez:2018uz,Mariani:2017jd,Ouni:2017db,Bavota:2014kr,CORTELLESSA2021106568}.
These techniques are especially effective on problems whose objectives can be expressed through quantifiable metrics.
Problems related to non-functional aspects undoubtedly fit into this category, as witnessed by the vast literature in this domain~\cite{DBLP:conf/mompes/AletiBGM09,Martens:2010bn}.
Most approaches are based on evolutionary algorithms~\cite{DBLP:journals/csur/BlumR03} that allow exploring the solution space by combining solutions.

One of the main drawbacks of applying optimization techniques to improve non-functional attributes is that, more often than not, the search for better alternatives requires a considerable amount of resources, notably time.
In fact, every time a new solution is generated, the algorithms usually have to quantify non-functional indices by solving non-functional models, either analytically or by simulating them.
Due to their complexity, it is difficult to further improve the efficiency of these activities.
Therefore, they intrinsically extend the time required for the search to obtain better solutions.

When performed on realistic models, this type of non-functional optimization can even take days~\cite{SEAA2021}.
This clearly poses an obstacle on the adoption of these techniques in any practical design and development scenario.
An alternative approach to the problem could be the imposition of a time limit on the search for better solutions.
Usually, this time cap is called the search budget, and it represents the maximum budget that can be spent to explore the solution space.
Hence, on the one hand, a smaller budget might heavily limit the exploration of the solution space, hampering the quality of the resulting Pareto fronts (\ie the set of non-dominated solutions obtained at the end of the optimization).
On the other hand, a time cap that is too large might defeat the purpose of having one in place~\cite{Arcuri_Fraser_2011}.

In our recent work~\cite{SEAA2021,CORTELLESSA2021106568,Arcelli:2018vo}, we experienced that the required time to investigate the solution space for model-based multi-objective refactoring optimization represents one of the main limitations.
Therefore, in this paper, we present an initial investigation on the influence of the search budget on the quality of solutions in the modeling context.
We show how a designer can find and evaluate a possible trade-off between the time spent on the search and the quality of solutions.
We also provide insights on the role of different searching policies, represented by different genetic algorithms, when imposing a time cap.
Specifically, we run experiments with three increasing time budgets and three different genetic algorithms.

To estimate the differences in the quality of Pareto fronts when varying the budget and the algorithm, we employ the Hypervolume (HV) quality indicator for multi-objective problems~\cite{BEUME20071653,Cao:2015dm}, and hypothesis testing.
The HV measures the amount of volume in the solution space that is covered by a computed Pareto front with respect to a reference Pareto front.
In our case, the reference Pareto front is one obtained without a time budget, but terminated after 102 evolutions.

We experiment on two model-based benchmarks, namely Train Ticket Booking Service~\cite{DBLP:conf/staf/Pompeo0CE19}, and CoCoME~\cite{cocome}.
Furthermore, we compare three genetic algorithms, \ie \nsga~\cite{Deb:2002ut}, \spea~\cite{zitzler2001spea2}, and \pesa~\cite{Corne_Jerram_Knowles_Oates_2001}, in order to identify which algorithm performs better when the search is limited in time.

This study answers the following research questions:

\begin{itemize}
 \item \textbf{RQ1}: To what extent does the time budget penalize the quality of Pareto fronts?
 \item \textbf{RQ2}: Which algorithm performs better when limited by a time budget?
\end{itemize}

Our results show that the time budget heavily impacts the quality of Pareto fronts.
Furthermore, we notice that slightly increasing the budget generates little improvements of Pareto fronts quality.
On the contrary, the choice of the algorithm seems to be crucial.
In most cases, \nsga is the fastest among the analyzed ones, while \pesa is the algorithm that generates the solutions with the highest quality.
Also, \spea shows worse performance than \nsga and \pesa.
Our findings suggest that, when in need for a faster algorithm, \nsga should be preferred, while \pesa can deliver better solutions in longer, but still reasonable, time.

The remaining of the paper is structured as follows: \secref{sec:related} reports related work, \secref{sec:background} introduces background concepts, \secref{sec:approach} presents the design of this study. 
Results are presented and discussed in \secref{sec:results}. 
\secref{sec:lesson} describes our findings towards the search budget in model-based multi-objective refactoring optimization. \secref{sec:conclusion} ends the paper and reports future work.
 \section{Related Work}\label{sec:related}

In the last decade, software model multi-objective optimization studies have been introduced to optimize various quality attributes (\eg reliability, and energy \cite{Martens:2010bn,5949650,DBLP:conf/qosa/MeedeniyaBAG10,10.1007/978-3-642-13821-8_8,CORTELLESSA2021106568}) with different degrees of freedom in the modification of models (\eg service selection~\cite{Cardellini:2009:QRA:1595696.1595718}).

Recent work compares the ability of two different multi-objective optimization approaches to improve non-functional attributes~\cite{NI2021106565,10.1145/3132498.3132509} within a specific modeling notation (\ie Palladio Component Model~\cite{Becker:2009cl}). 
To find optimal solutions, the authors apply architectural tactics, which mainly change system configurations (\eg hardware settings, or operation demands).
Conversely, in this work, we apply refactoring actions that change the structure of the initial model by preserving the original behavior.
Another difference is the modeling notation, as we use UML with the goal of experimenting on a standard notation instead of a custom Domain Specific Language.

Menasce~\etal have presented a framework for architectural design and quality optimization~\cite{DBLP:conf/wosp/MenasceEGMS10}, where architectural patterns are used to support the searching process (\eg load balancing, fault tolerance). 
Two limitations affects the approach: the architecture has to be designed in a tool-specific notation and not in a standard modeling language (as we do in this paper), and performance indices are computed through equation-based analytical models that could be too simple to capture architectural details and resource contention.

Aleti \etal~\cite{DBLP:conf/mompes/AletiBGM09} have presented an approach for modeling and analyzing AADL architectures~\cite{DBLP:books/daglib/0030032}. 
They have also introduced a tool aimed at optimizing different quality attributes while varying the architecture deployment and the component redundancy.
Our work relies on UML models and considers more complex refactoring actions, as well as different target attributes for the fitness function.
Besides, we investigate the role of performance antipatterns in the context of many-objective software model refactoring optimization.

 \section{Background}\label{sec:background}

\begin{figure}
	\centering
	\includegraphics[width=.9\linewidth]{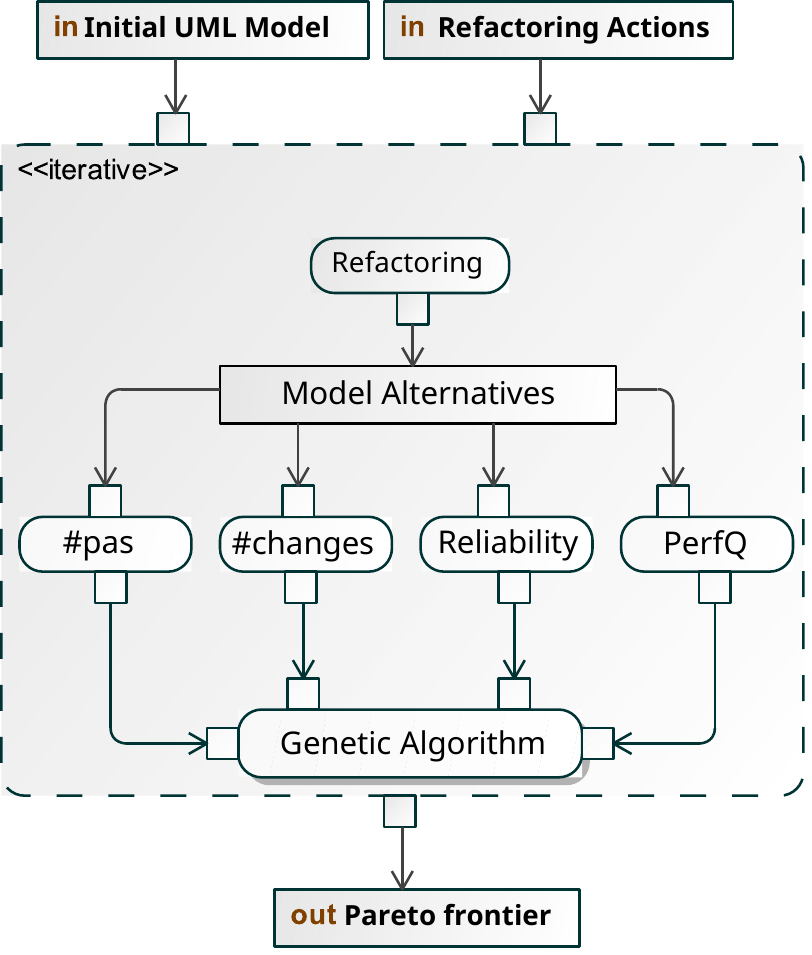}
	\caption{\label{fig:approach}The graphical representation of the approach. The approach takes as input: the set of all the available refactoring actions (\emph{Refactoring Actions}), and the subject model (\emph{Initial UML Model}). The \emph{Genetic Algorithm} (\ie \nsga, \spea, and \pesa) randomly selects and combines refactoring actions (\emph{Refactoring}) in order to build a set of \emph{Model Alternatives}. \emph{\pas}, \emph{\achanges}, \emph{\reliability}, and \emph{\perfq} are the four objectives that drive the optimization process.}
\end{figure}

In this study we analyzed the impact of search budget on three \emph{Genetic Algorithms}: \nsga~\cite{Deb:2002ut}, \spea~\cite{zitzler2001spea2}, and \pesa~\cite{Corne_Jerram_Knowles_Oates_2001}. 
We chose these algorithms for their different policies in searching the solution space. 
For example, \nsga uses the knowledge of non-dominated sorting to generate Pareto frontiers, \spea uses two archives to store computed Pareto frontiers, and \pesa uses the hyper-grid concept to compute Pareto frontiers.
Our process, depicted in \figref{fig:approach}, optimizes four conflicting objectives: the performance overall quality indicator (\perfq)~\cite{Arcelli:2018vo}, the reliability (\reliability) of model alternatives~\cite{DBLP:journals/infsof/CortellessaET20}, the number of performance antipatterns(\pas), and the architectural distance (\achanges)~\cite{SEAA2021}.

\paragraph*{Performance Quality Indicator (\perfq)}\label{sec:background:perfq}

\perfq quantifies the performance improvement/detriment between two models.
Furthermore, a single \perfq for each performance index is computed as the normalized ratio between the index value of a model alternative and the initial model.
Finally, the global \perfq is computed as the average across the number of performance indices considered in the performance analysis.

\paragraph*{Reliability model}\label{sec:background:reliability}

The reliability model that we adopt here to quantify the \reliability objective is based on the model introduced in~\cite{CortellessaSC02}.

The model takes into account failure probabilities of components and communication links, as well as the probability of a scenario to be executed.
Such probabilities are combined to obtain the overall reliability on demand of the system, which represents how often the system is not expected to fail when its scenarios are invoked.

\paragraph*{Performance Antipatterns}\label{sec:background:pas}

A performance antipattern describes bad design practices that might lead to performance degradation in a system. These textual descriptions were later translated into first-order logic (FOL) equations~\cite{DBLP:conf/fase/ArcelliCT15}.

FOLs enable an automated comparison with thresholds in order to reveal the occurrences of a performance antipattern.
The identification of such thresholds is a non-trivial task, and using deterministic values may result in an excessively strict detection, where the smallest change in the value of a literal determines the occurrence of the antipattern.
For these reasons, we use the fuzzy threshold concept~\cite{DBLP:conf/fase/ArcelliCT15}, instead of detecting a performance antipattern in a deterministic way.
By using fuzzy thresholds, we assign probabilities to the occurrences of antipatterns. 

\paragraph*{Architectural distance}\label{sec:background:distance}

The architectural distance \achanges quantifies the distance of the model obtained by applying refactoring actions to the initial one. 
The effort needed to perform a refactoring is quantified as the product between the \emph{baseline refactoring factor}, which is associated to each refactoring action, and the \emph{architectural weight}, which is associated to each model element on the basis of the number of connections to other elements in the model~\cite{SEAA2021}. 
The overall \achanges is obtained by summing the efforts of all refactoring actions contained in a solution.

\paragraph*{Hypervolume}\label{sec:background:hv}

Establishing the quality of a computed Pareto front is arduous, and it is a NP-hard problem~\cite{Ali_Arcaini_Pradhan_Safdar_Yue_2020}.
Different quality estimators have been introduced, such as the Hypervolume (HV)~\cite{Cao:2015dm,BEUME20071653}.
Each estimator measures a different quality aspect of a Pareto front.
In this study, we use the HV as our quality estimator, since it has been proved to be a valid estimator for Pareto fronts comparison~\cite{Falcon-Cardona_Coello_2021}.
Moreover, the HV interpretation is straightforward in the context of our problem.

The HV measures the amount of the volume of the solution space that a Pareto front (\computedP) covers with respect to a reference Pareto front (\referenceP), and it can assume values between $0$ and $1$.
When the $HV = 0$, it means that the \computedP is fully dominated by the \referenceP, while $HV = 1$ means that each point within the \computedP is non-dominated by any points within the \referenceP.
Therefore, the closer to 1 the HV, the higher the quality of the \computedP.
In our evaluation, we use the HV to estimate the quality of the \computedP obtained with a search budget when compared to a \referenceP computed without, but terminated after 102 evolutions.
 \section{Study Design}\label{sec:approach}

The goal of the study is to establish how much the imposition of a time-based search budget can hamper the quality of the resulting Pareto fronts in a model-based multi-objective refactoring optimization context.
Additionally, we are interested in how different algorithms cope with different search budgets.
To this extent, we selected two case studies, and we run the optimization with search budgets of \texttt{15}, \texttt{30}, and \texttt{60} minutes.
Moreover, for each search budget, we also run three genetic algorithms: \nsga, \spea, and \pesa.
We chose these three algorithms on the basis of their different searching policies, as described in~\secref{sec:background}.

As recommended in other studies~\cite{Zitzler_Deb_Thiele_2000}, because of the random nature of genetic algorithms, we run the same experiment 31 times, and we compute the HV for each computed Pareto front (\computedP).
The HV indicator measures the amount of volume in the solution space that is covered by a \computedP with respect to an optimal reference Pareto front (\referenceP) for the problem.
Since \referenceP is unknown in our case studies, we computed the HV with respect to the best Pareto front we obtained for the same case studies when we run the search for 102 evolutions without a search budget. 
The entire study consisted of \textbf{558} experiments\footnote{The replication package: \url{https://zenodo.org/record/6446516}} that we performed on three AMD EPYC 7282, each with 64 cores and 512GB of RAM.

We follow the guidelines by Arcuri~\etal~\cite{ArcuriB14} to compare the experiments against each other.
Therefore, we apply the \mwu non-parametric statistical test (also referred to as Wilcoxon rank-sum test)~\cite{MWU} with the null hypothesis that the experiments do not show a statistically significant difference.
We consider two experiments to be significantly different on the basis of their HV if the test computes a p-values smaller than $0.05$.
To assess the magnitude of the difference, we use the Vargha--Delaney \vda~\cite{VDA}, a standardized non-parametric effect size measure.
\vda takes values between $0$ and $1$, and a values of $0.5$ indicates that the two experiments are equivalent.
The closer the value of \vda gets to $0$ or $1$, the larger the effect.
The interpretation of the magnitude as negligible, small, medium, and large is performed according to the thresholds $0.147$, $0.33$, $0.474$~\cite{HessKromrey2004}.

\section{Empirical Evaluation}\label{sec:results}

In this section, we present the results of our experiments, and we answer the research questions formulated in \secref{sec:intro}.

\subsection{Case Study}\label{sec:case-study}

We applied our optimization approach to two case studies from the literature: the Train Ticket Booking Service (\ttbs)~\cite{DBLP:conf/staf/Pompeo0CE19}, and the well-established modeling case study \ccm, whose UML model has been derived by the specification in~\cite{cocome}.\footnote{\url{https://github.com/SEALABQualityGroup/uml2lqn-casestudies}}

\medskip

\paragraph*{Train Ticket Booking Service}

Train Ticket Booking Service (\ttbs) is a web-based booking application, whose architecture is based on the microservice paradigm.
The system is made up of 40 microservices, and it provides different scenarios through users that can perform realistic operations, \eg book a ticket or watch trip information.

For our analysis we downsized the \ttbs UML model~\cite{DBLP:conf/staf/Pompeo0CE19} by considering \expComp{11}, \expNode{11}, and \expUC{3}.
Furthermore, we considered \emph{Login}, \emph{Update user details} and \emph{Rebook} as selected scenarios because they commonly represent performance-critical scenarios in a ticketing booking service.
Also, the model defines two user categories: simple and admin users.

\medskip

\paragraph*{\ccm}

\ccm describes a trading system containing several stores.
A store can have one or more cash desks for processing goods.
A cash desk is equipped with all the tools needed to serve a customer (\eg a Cash Box, Printer, Bar Code Scanner). 
\ccm covers possible scenarios performed at a cash desk (\eg scanning products, paying by credit card, or ordering new goodies).
\ccm describes 8 scenarios involving more than 20 components.

For our analysis, we downsized \ccm by selecting \expUC{3}, \expComp{13}, and \expNode{8}.
Beside this, we focused on three scenarios: \emph{Process Sale}, \emph{Receive Ordered Products}, and \emph{Show stock reports} because they represent common activities in a trading system.

\subsection{RQ1}\label{sec:rq1}

\RQ{RQ1}{To what extent does the time budget penalize the quality of Pareto fronts?}

The first main concern on the imposition of a search budget is the effect this can have on the optimization process.
As outlined in \secref{sec:approach}, we estimate the impact a search budget has on the optimization by means of the HV quality indicator.
\tabref{tab:qi_indicators} reports, for each algorithm and for each search budget, the average HV achieved in \independentRun runs, along with its standard deviation.
These value represent the percentage of volume of the \referenceP that is covered by a give \computedP.
Intuitively, this gives an idea of how much of the solution space was covered with the budget restriction, compared to a run without the search budget.
We can observe that, in fact, the time budget heavily impacts the quality of the obtained Pareto fronts.

\begin{table}
\centering
\begin{tabular}{lccr}
\toprule
Algor. &  Budget & HV avg &     HV stdev \\
\midrule
\multicolumn{4}{c}{\ttbs ($\Omega = 1.2 \times 10^{13}$)}\\
\midrule

\nsga & 15 min &  0.3060 &    0.0915 \\
\nsga & 30 min &  0.3469 &    0.1071 \\
\nsga & 60 min &  0.3437 &    0.0980 \\
\pesa & 15 min &  0.3532 &    0.0794 \\
\pesa & 30 min &  0.4084 &    0.0757 \\
\pesa & 60 min &  0.4182 &    0.0819 \\
\spea & 15 min &  0.3041 &    0.0794 \\
\spea & 30 min &  0.2917 &    0.0920 \\
\spea & 60 min &  0.2868 &    0.0769 \\

\midrule
\multicolumn{4}{c}{\ccm ($\Omega = 3.26 \times 10^{16}$)}\\
\midrule

\nsga & 15 min &  0.0931 &    0.0335 \\
\nsga & 30 min &  0.1199 &    0.0523 \\
\nsga & 60 min &  0.1125 &    0.0604 \\
\pesa & 15 min &  0.1363 &    0.0277 \\
\pesa & 30 min &  0.1460 &    0.0300 \\
\pesa & 60 min &  0.1514 &    0.0366 \\
\spea & 15 min &  0.1189 &    0.0336 \\
\spea & 30 min &  0.1098 &    0.0309 \\
\spea & 60 min &  0.1023 &    0.0384 \\

\bottomrule
\end{tabular}
\caption{Average HV quality indicator and its standard deviation over 31 runs, listed by algorithm and search budget. 
Higher values are associated to a better quality of the Pareto fronts.
$\Omega$ is the size of the solution space computed as the Cartesian product of the types of refactoring actions and all the eligible refactoring targets in any possible refactoring sequence.}
\label{tab:qi_indicators}
\end{table}
 
The search budget had a different impact on the two case studies.
In \ttbs, the search was able to achieve better HV in all cases, when compared to \ccm.
This is probably due to the difference in size and complexity between the two case studies.
\ccm not only has a larger number of possible refactoring candidates, but its model defines a more complex behavior.
This inherently leads to a bigger solution space ($\Omega$ in the table), but also to spending more time in computing the objective functions.
Therefore, on average, the longer it takes to complete a single evolution, the fewer the  evolutions will be performed on a given time budget.

To assess whether doubling or quadrupling the time budget makes a significant difference in the HV of the \computedP, we compare the results obtained with different budgets but with the same algorithm.
\tabref{tab:hv_test_time} reports the results of the \mwu test, and the corresponding \vda effect size.
The p-value is highlighted in bold when the detected difference is statistically significant.
The time budget is underlined when (i) the test resulted in a significant difference, and (ii) the experiment running on that time budget produced higher values of HV.
In very few cases (two per case study), we obtained a significant difference, and in all the cases this was detected for the \pesa algorithm: with a medium magnitude in \ttbs, and with a large one in \ccm.
This suggests that, except for \pesa, the main difference in the obtained HV values might be imputed to a difference in the used algorithm, more than to a difference in the budget.
We are going to investigate this in the next section.

\begin{table}[]
\centering
\begin{tabular}{llllll}
    
\toprule
Algor. & Budget 1 & Budget 2 &              MWU p &                      \vda & \\
\midrule
\multicolumn{6}{c}{\ttbs} \\
\midrule
\nsga &   15 min &   30 min &          0.1677 & (S) \ebar{0.3975}{0.1025} \\
\nsga &   15 min &   60 min &          0.1677 & (S) \ebar{0.3975}{0.1025} \\
\nsga &   30 min &   60 min &          0.9327 & (N) \ebar{0.5068}{0.0068} \\
\pesa &   15 min &   \underline{30 min} &  \textbf{0.018} & (M) \ebar{0.3247}{0.1753} \\
\pesa &   15 min &   \underline{60 min} & \textbf{0.0031} &   (M) \ebar{0.281}{0.219} \\
\pesa &   30 min &   60 min &          0.4223 & (N) \ebar{0.4402}{0.0598} \\
\spea &   15 min &   60 min &          0.4992 & (N) \ebar{0.5505}{0.0505} \\
\spea &   30 min &   15 min &          0.4556 & (N) \ebar{0.4443}{0.0557} \\
\spea &   30 min &   60 min &          0.7999 & (N) \ebar{0.4807}{0.0193} \\
\midrule
\multicolumn{6}{c}{\ccm} \\
\midrule
\nsga &   15 min &   30 min &             0.0574 &   (S) \ebar{0.359}{0.141} \\
\nsga &   60 min &   15 min &             0.1054 & (S) \ebar{0.6202}{0.1202} \\
\nsga &   60 min &   30 min &             0.8769 &   (N) \ebar{0.488}{0.012} \\
\pesa &   15 min &   \underline{30 min} & \textbf{$<$0.0001} & (L) \ebar{0.2092}{0.2908} \\
\pesa &   \underline{60 min} &   15 min & \textbf{$<$0.0001} & (L) \ebar{0.7992}{0.2992} \\
\pesa &   60 min &   30 min &             0.6024 &   (N) \ebar{0.539}{0.039} \\
\spea &   30 min &   15 min &             0.2483 & (S) \ebar{0.4142}{0.0858} \\
\spea &   60 min &   15 min &             0.1249 & (S) \ebar{0.3861}{0.1139} \\
\spea &   60 min &   30 min &             0.6123 &   (N) \ebar{0.462}{0.038} \\
\bottomrule

\end{tabular}
\caption{\mwu test and \vda effect sizes comparing the HV achieved with different time budgets in 31 runs. Magnitude interpretation: negligible (N), small (S), medium (M), large (L). The magnitude of the effect size is also represented by bars.}
\label{tab:hv_test_time}
\end{table}

\subsection{RQ2}\label{sec:rq2}

\RQ{RQ2}{Which algorithm performs better when limited by a time budget?}

When a time constraint is imposed on the process, a designer may be interested in selecting the algorithm that provides best quality solutions for the specific time budget.
In this section, we discuss some aspects on which such a decision could be based.

At first, we compare the algorithms against each other on the basis of the HV they achieved in the experiments, analogously to how it has been done in \secref{sec:rq1}.
\tabref{tab:hv_test_algo} reports the results of the \mwu test, and the corresponding \vda effect size.
The name of the algorithm is underlined when (i) the test resulted in a significant difference, and (ii) that algorithm yielded higher values of HV.
In this case, most of the tests revealed a significant difference between the algorithms in any given time budget (highlighted in bold).
\pesa performed better in many cases and in both case studies, \nsga scored better on only two cases in \ttbs and not by a large margin, and \spea won only the 15 minutes budget test in \ccm.

\begin{table}[]
\centering
\begin{tabular}{llllll}
    
\toprule
Budget & Algor. 1 & Algor. 2 &              MWU p &               \vda & \\
\midrule
\multicolumn{6}{c}{\ttbs} \\
\midrule
15 min &    \underline{\pesa} &    \nsga &    \textbf{0.0487} & (S) \ebar{0.6462}{0.1462} \\
15 min &    \spea &    \nsga &             0.8548 &   (N) \ebar{0.486}{0.014} \\
15 min &    \spea &    \underline{\pesa} &    \textbf{0.0234} & (M) \ebar{0.3319}{0.1681} \\
30 min &    \nsga &    \underline{\pesa} &    \textbf{0.0167} & (M) \ebar{0.3226}{0.1774} \\
30 min &    \spea &    \underline{\nsga} &    \textbf{0.0385} & (S) \ebar{0.3465}{0.1535} \\
30 min &    \spea &    \underline{\pesa} & \textbf{$<$0.0001} & (L) \ebar{0.1582}{0.3418} \\
60 min &    \nsga &    \underline{\pesa} &    \textbf{0.0037} & (M) \ebar{0.2851}{0.2149} \\
60 min &    \spea &    \underline{\nsga} &    \textbf{0.0202} & (M) \ebar{0.3278}{0.1722} \\
60 min &    \spea &    \underline{\pesa} & \textbf{$<$0.0001} & (L) \ebar{0.1301}{0.3699} \\
\midrule
\multicolumn{6}{c}{\ccm} \\
\midrule
15 min &    \nsga &    \underline{\spea} &    \textbf{0.0085} & (M) \ebar{0.3049}{0.1951} \\
15 min &    \underline{\pesa} &    \nsga &    \textbf{0.0072} & (M) \ebar{0.6993}{0.1993} \\
15 min &    \pesa &    \spea &             0.7999 & (N) \ebar{0.4807}{0.0193} \\
30 min &    \underline{\pesa} &    \nsga &    \textbf{0.0066} & (M) \ebar{0.7014}{0.2014} \\
30 min &    \spea &    \nsga &             0.5543 & (N) \ebar{0.4558}{0.0442} \\
30 min &    \spea &    \underline{\pesa} & \textbf{$<$0.0001} & (L) \ebar{0.1738}{0.3262} \\
60 min &    \underline{\pesa} &    \nsga &    \textbf{0.0127} & (M) \ebar{0.6847}{0.1847} \\
60 min &    \spea &    \nsga &             0.3789 & (N) \ebar{0.4344}{0.0656} \\
60 min &    \spea &    \underline{\pesa} & \textbf{$<$0.0001} & (L) \ebar{0.1686}{0.3314} \\
\bottomrule

\end{tabular}
\caption{\mwu test and \vda effect sizes comparing the HV achieved by different algorithms in 31 runs. Magnitude interpretation: negligible (N), small (S), medium (M), large (L). The magnitude of the effect size is also represented by bars.}
\label{tab:hv_test_algo}
\end{table}
 
\begin{figure*}\centering
   \begin{subfigure}{.98\linewidth}
      \includegraphics[width=.96\textwidth]{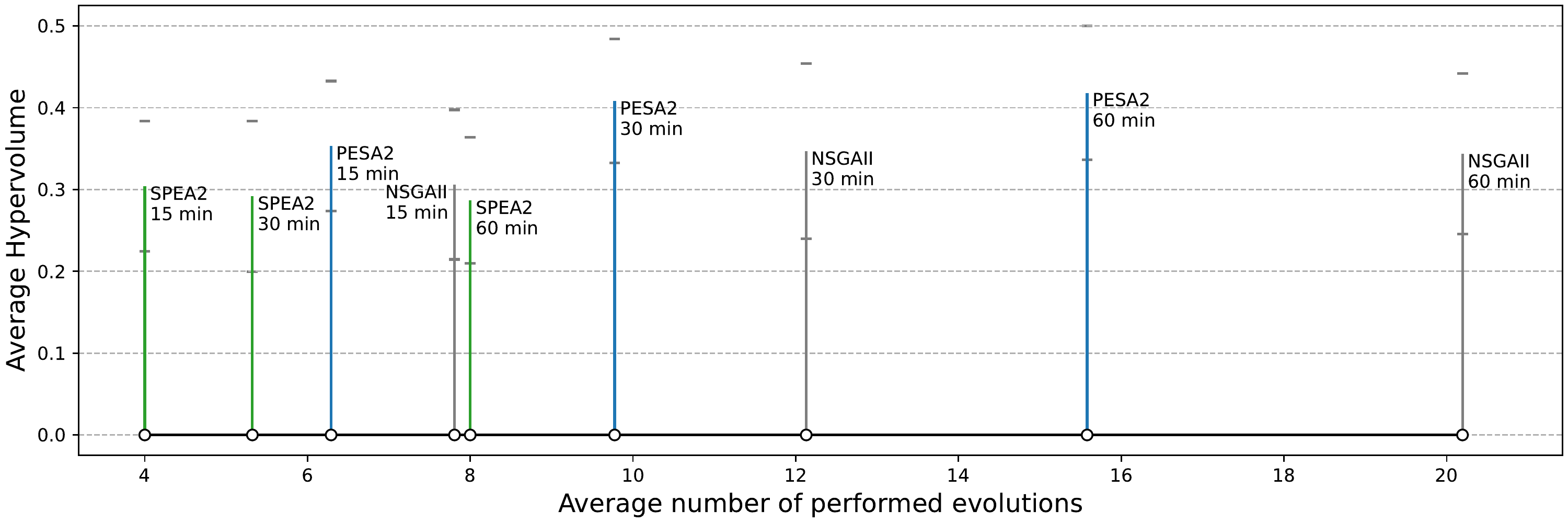}
      \caption{\ttbs}
      \label{fig:ttbs_HV_timeline}
   \end{subfigure}\hfill
\begin{subfigure}{.98\linewidth}
      \includegraphics[width=.96\textwidth]{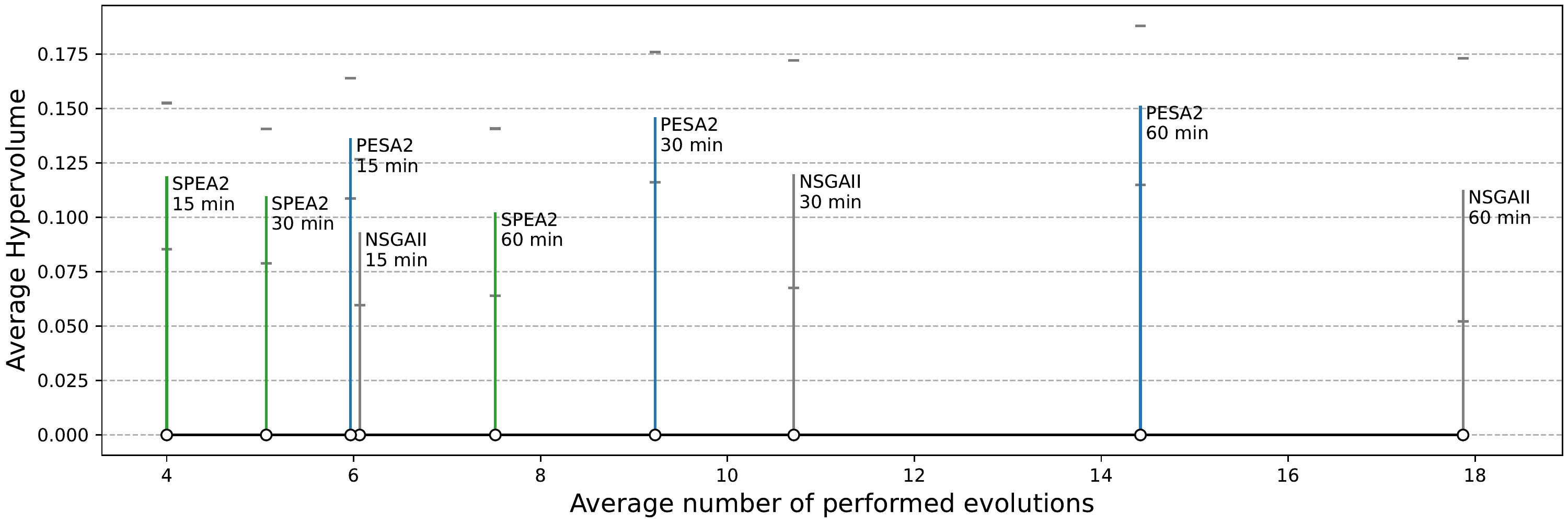}
      \caption{\ccm}
      \label{fig:simplified-cocome_HV_timeline}
   \end{subfigure}\hfill
   \caption{Timelines of the number of evolutions performed by the algorithms in the different budget configurations, along with the achieved HV. Vertical bars show the average HV over 31 runs, and ticks represent the standard deviation from the mean.}
\label{fig:hypervolume-timelines}
\end{figure*}

While the results vary in the two case studies we considered, the \pesa algorithm looks like a better choice in most cases.
To investigate the possible reasons behind such differences in HV, we take a look at how the HV is achieved and when, by comparing it to the time budget and the number of performed evolutions.
To this extent, \figref{fig:hypervolume-timelines} depicts the timelines of how the HV indicator varies with different search budgets, and how many evolutions were performed during the search.
From the timelines, we can see that \spea is the slowest algorithm in our experiments, whereas \nsga is the fastest one.
Furthermore, for each search budget, \nsga performed the highest number of evolutions, \eg it performed on average 20 evolutions for \ttbs with 60 minutes of search budget, and almost 18 for \ccm.
Conversely, \spea performed, on average, only 8 evolutions for \ttbs with a 60 minutes search budget. Concerning the \pesa algorithm, we can state that it consistently generated the highest HV in each case study and for every search budget.
However, it is slower than \nsga, but faster than \spea.

Analyzing the \ttbs results, we observed that the HV values of \spea almost lie close to $0.3$ fore every search budget, while the longer the search budget, the higher the HV values of \pesa.
The HV values of \nsga increase between the 15 and 30 minutes budget, while they are almost flat between 30 and 60 minutes.
In addition, we can observe that the timelines of the two case studies resemble each other.
In fact, also per \ccm, \nsga is the fastest algorithm, \spea the slowest, and \pesa generates the highest HV values.
Furthermore, the number of evolutions are consistent with the number of evolutions of \ttbs. 

\begin{figure*}
   \begin{subfigure}{\dimexpr0.30\textwidth+20pt\relax}\centering
     \makebox[20pt]{\raisebox{40pt}{\rotatebox[origin=l]{90}{\ttbs}}}\includegraphics[width=\dimexpr\linewidth-20pt\relax]{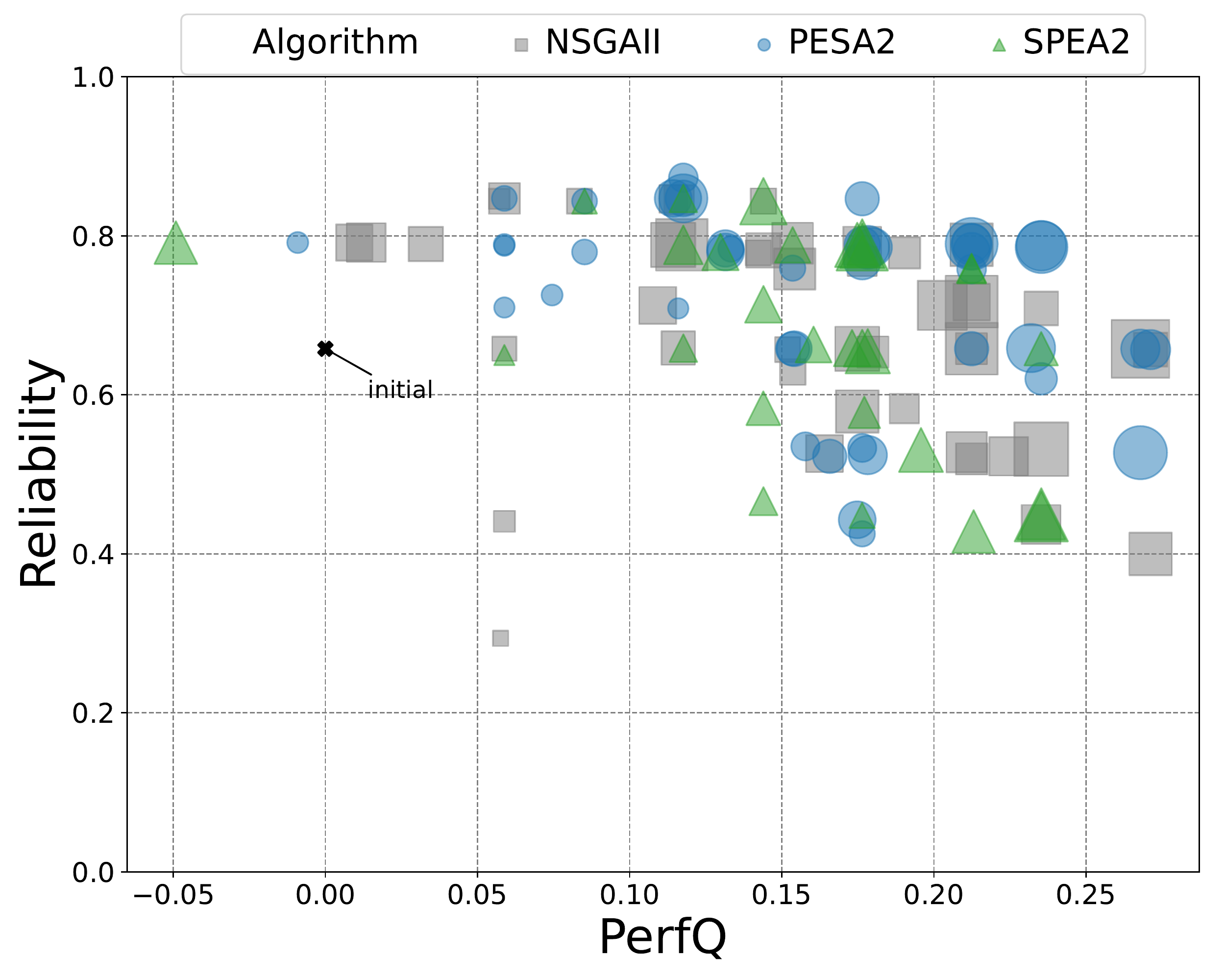}
     \makebox[20pt]{\raisebox{40pt}{\rotatebox[origin=l]{90}{\ccm}}}\includegraphics[width=\dimexpr\linewidth-20pt\relax]{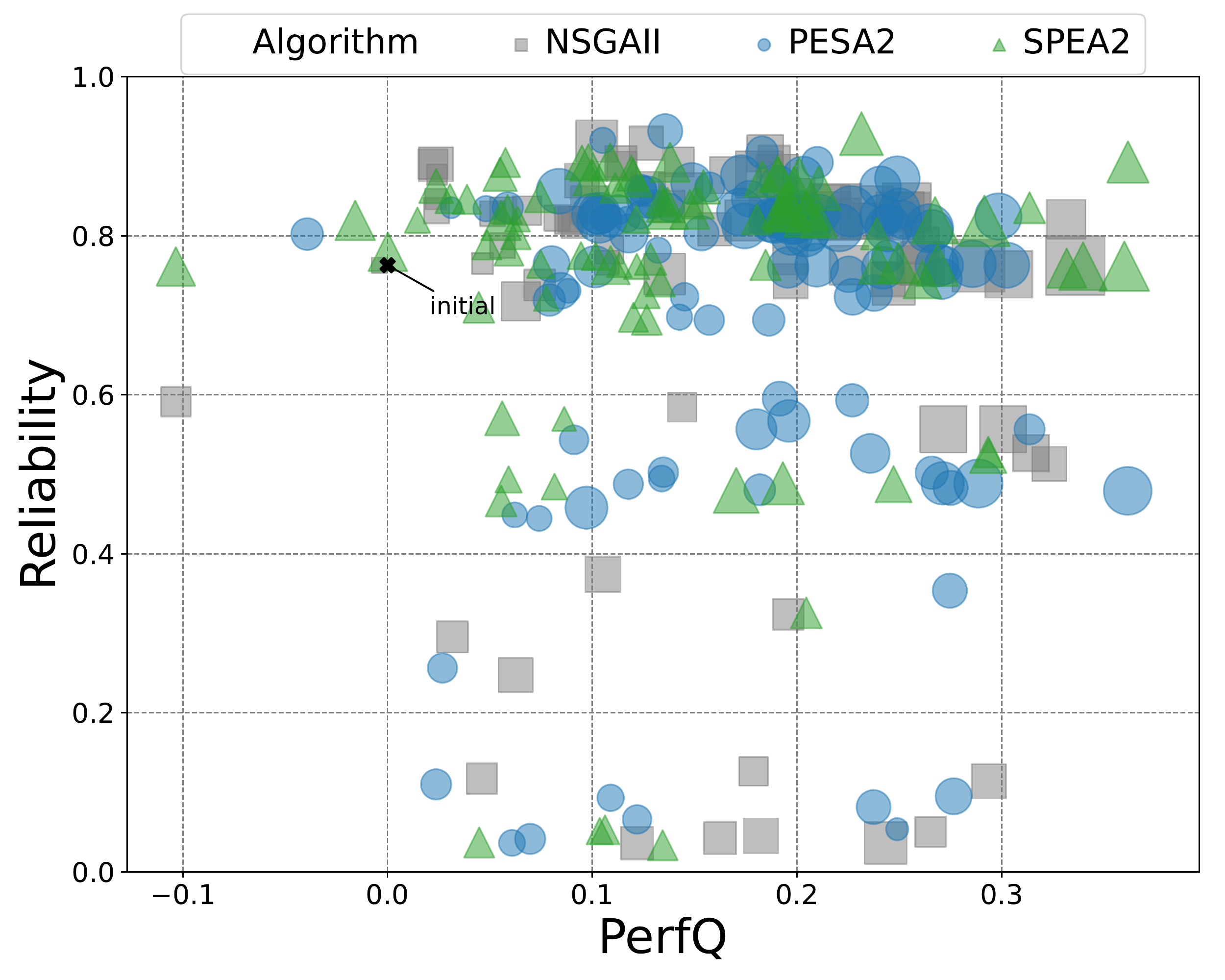}
     \caption{15 min budget}
     \label{fig:byalgo_15_min}
   \end{subfigure}\hfill \begin{subfigure}{.65\columnwidth}
     \centering
     \includegraphics[width=.94\linewidth]{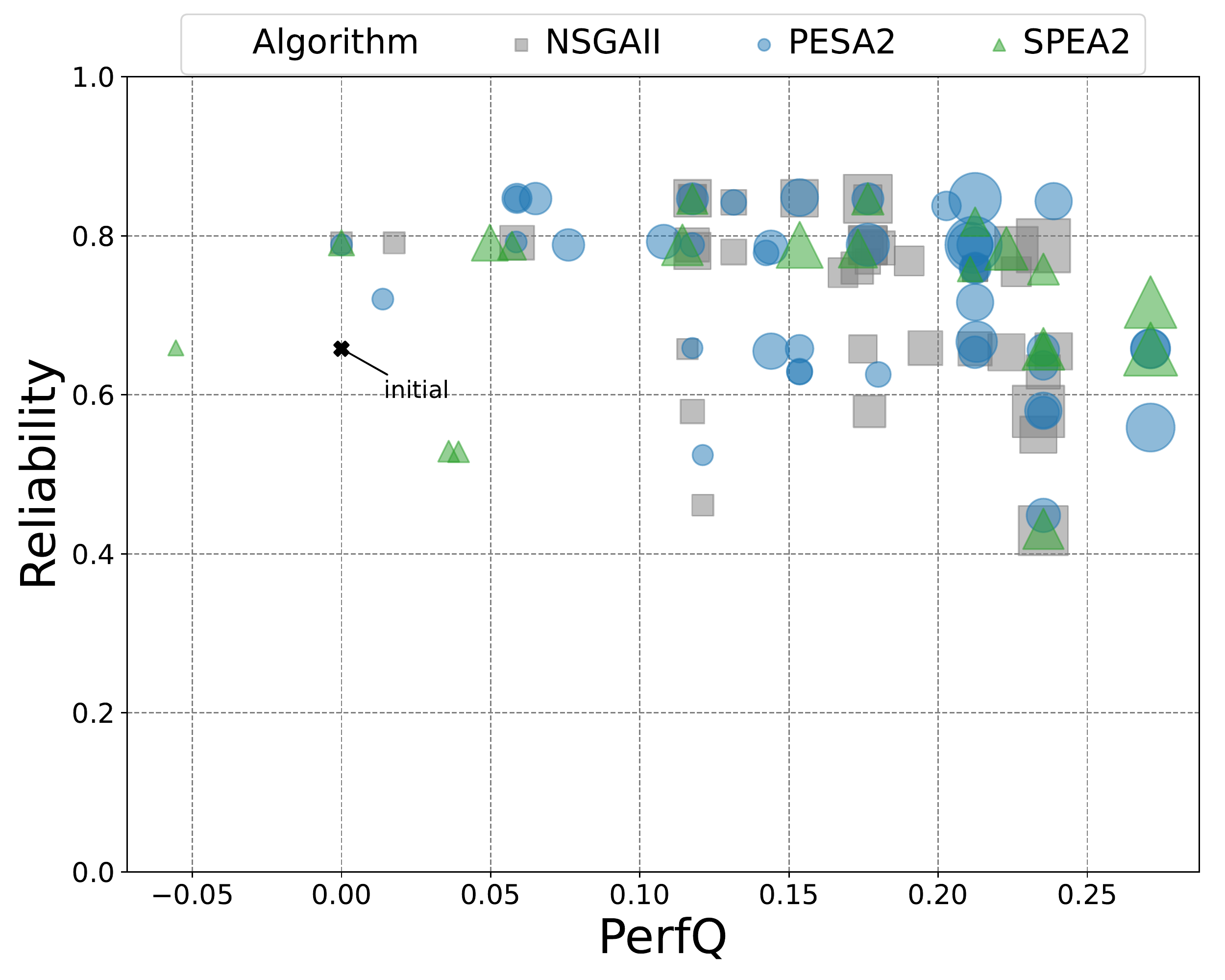}
     \includegraphics[width=.94\linewidth]{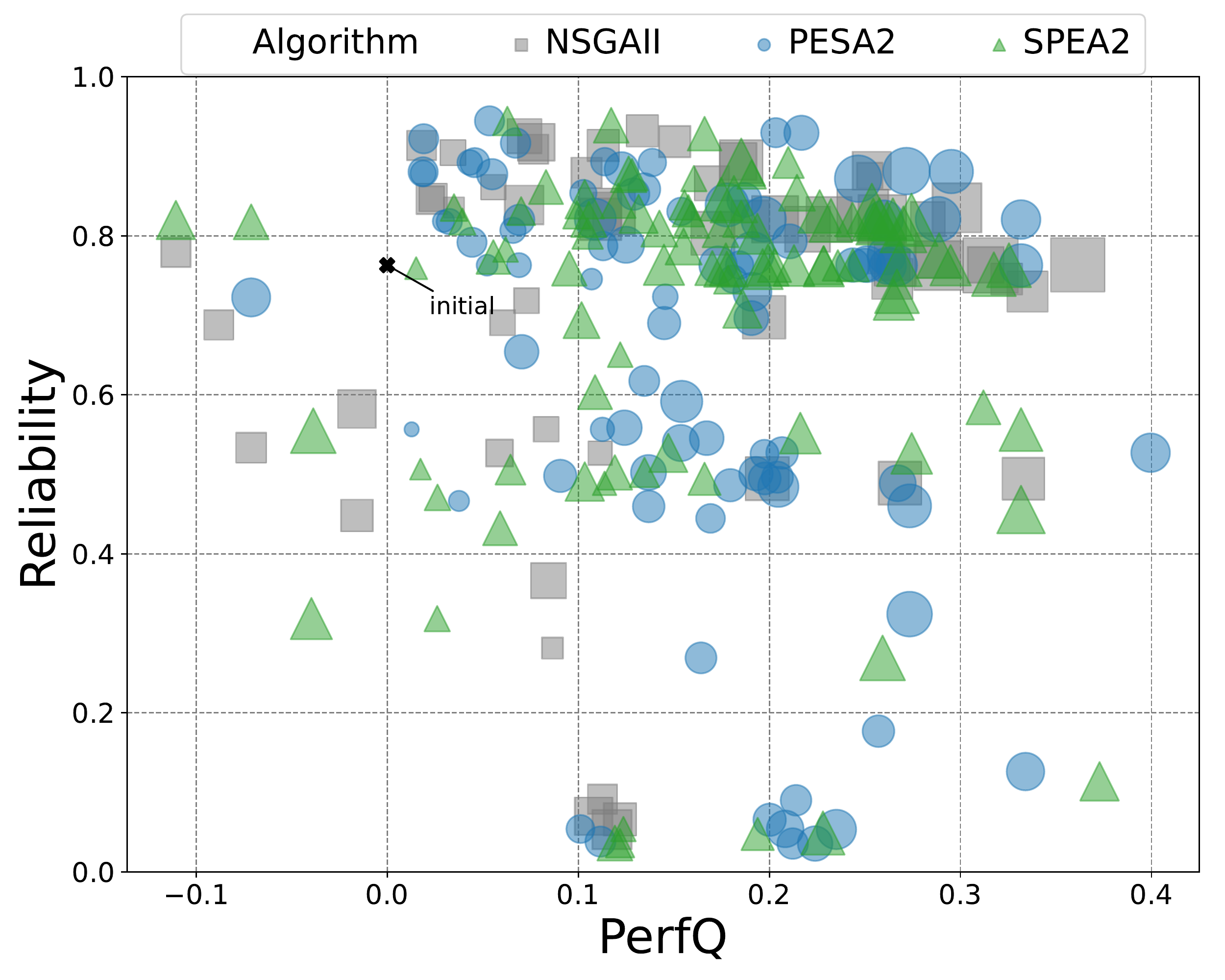}
     \caption{30 min budget}
     \label{fig:byalgo_30_min}
   \end{subfigure}\hfill \begin{subfigure}{.65\columnwidth}
     \centering
     \includegraphics[width=.94\linewidth]{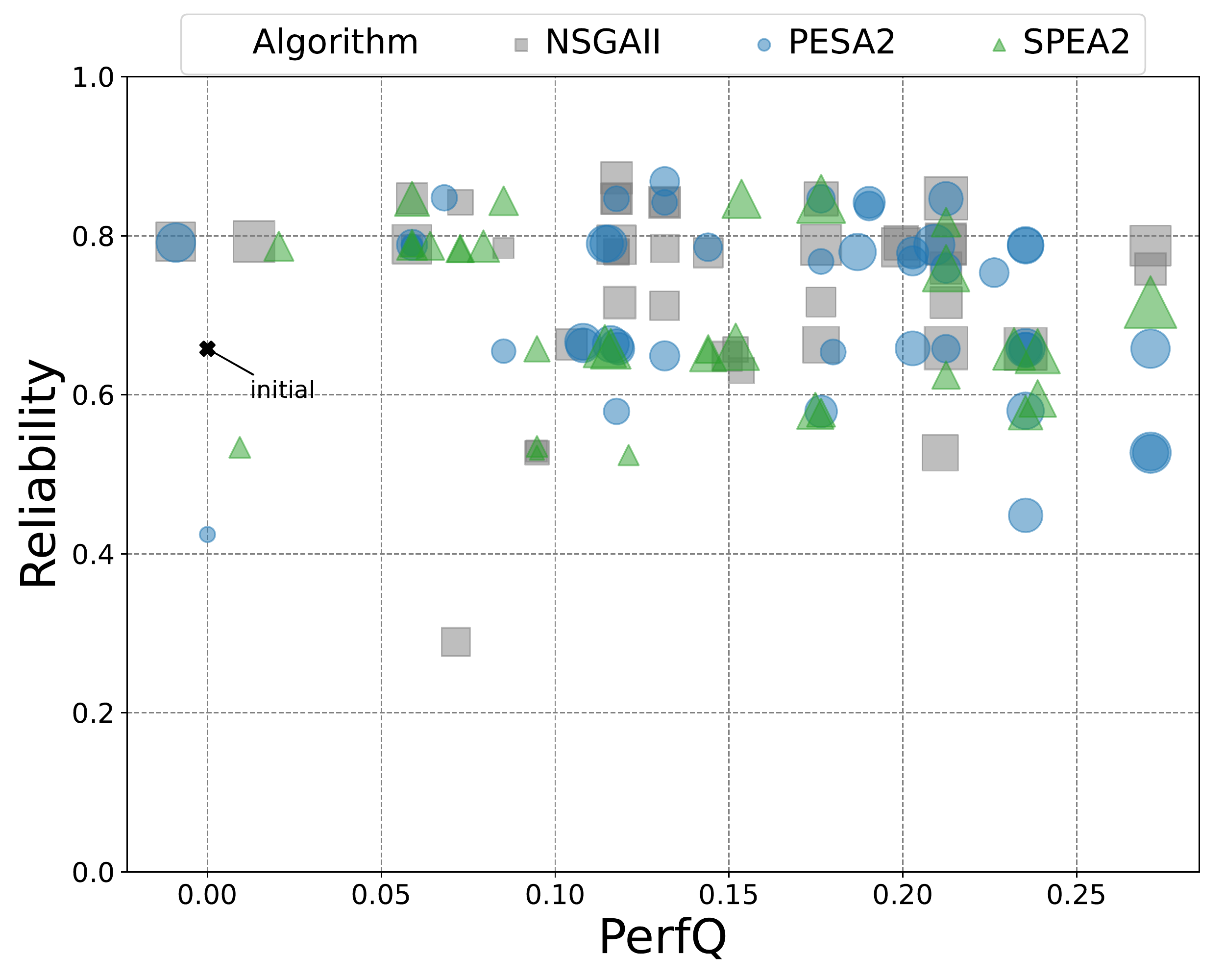}
     \includegraphics[width=.94\linewidth]{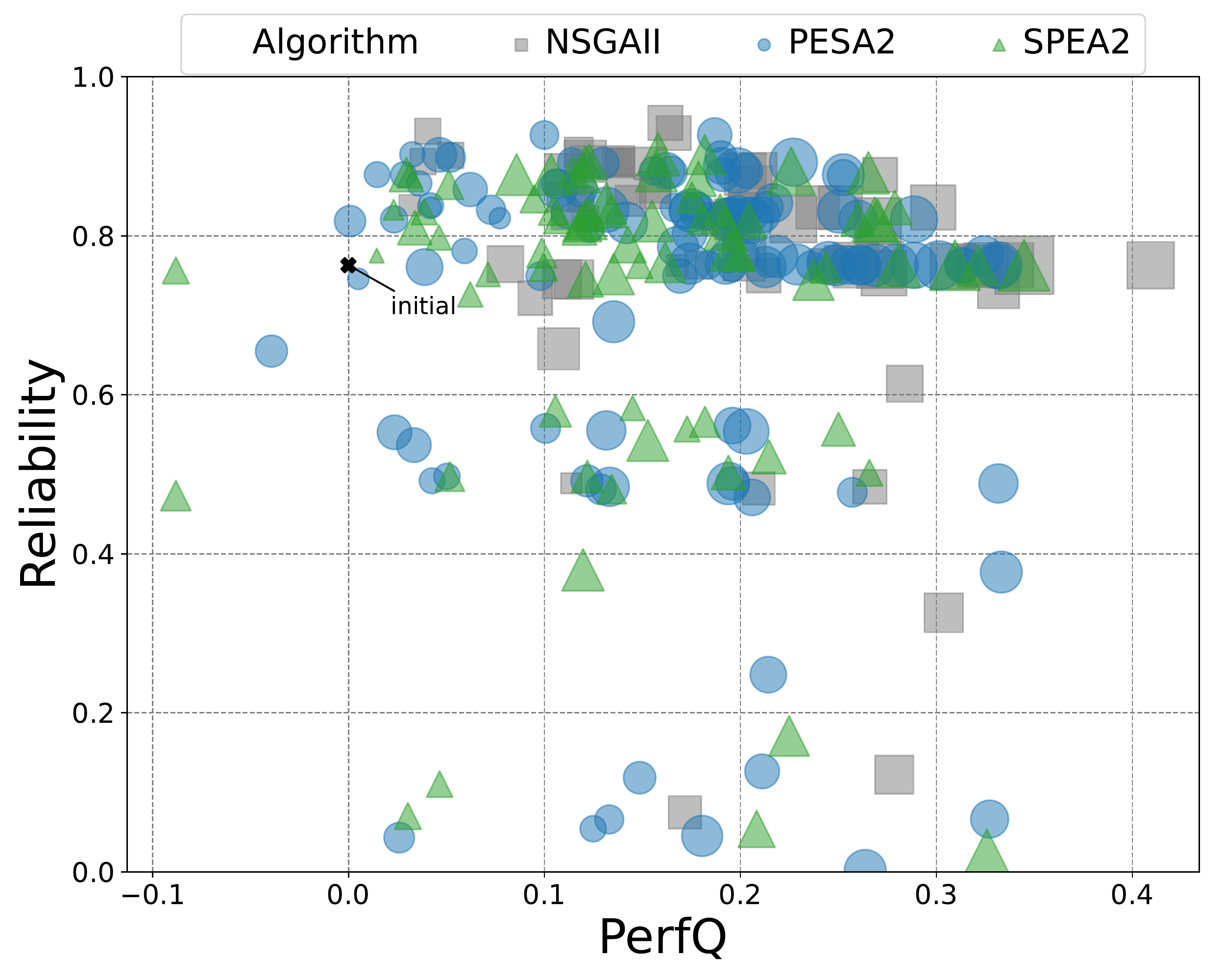}
     \caption{60 min budget}
     \label{fig:byalgo_60_min}
   \end{subfigure}\hfill \caption{\ttbs, and \ccm Pareto frontiers obtained by the three algorithms when varying the time budget between 15, 30, and 60 minutes. The top-right corner is the optimal point, whereas the bottom-left corner is the worst one. Filled symbols are the three algorithms: \nsga is the gray squares, \pesa is the blue circles, and \spea is the green triangles.}
  \label{fig:byalgo_scatter}
\end{figure*}

Another viewpoint on the difference among the algorithms could be the actual quality of the computed solutions in terms of the non-functional properties we are interested in.
To visually inspect this aspect, we produced scatter plots to compare \perfq and \reliability, because these objectives are the non-functional properties we aim at improving through the refactoring and optimization process.
Therefore, \figref{fig:byalgo_15_min}, \figref{fig:byalgo_30_min}, and \figref{fig:byalgo_60_min} depict the three \computedP when varying the time budget of all three genetic algorithms for both case studies.
At a glance, we can observe more densely populated \computedP for \ccm than for \ttbs, while \ttbs showed a more evident trend towards the top-right corner (the optimization direction for these two objectives).
Regarding the \ccm \computedP, we can observe an horizontal clustering for the three search budgets.
The cluster that lies around $0.8$ \reliability is always more populated that the other two: one between $0.4$ and $0.6$, and the other between $0.0$ and $0.2$, approximately.
We did not observe an evident motivation for the horizontal clustering of \ccm. 
We can only suppose that the characteristics of the \ccm model, which has a more complex behavior than \ttbs, prevent the algorithms from reaching higher \reliability values in the search budgets we considered.
Also, the \ccm solution space might be less homogeneous, with feasible solutions that are inherently clustered.

Summarizing, on the one hand we can establish a clear difference among the algorithms when comparing them on the basis of a quality indicator for multi-objective optimization, like the HV, and when looking at their speed in completing evolutions.
But on the other hand, if we only look at the non-functional properties we considered, there is not much difference in the shape of the \computedP and in the explored design space.

\subsection{Threats to validity}\label{sec:t2v}

In this section we discuss threats that might affect our results.

\paragraph*{Construct validity}
Our approach might be affected by \emph{Construct validity} threats.
An aspect that might affect our results is the estimation of the reference Pareto front (\referenceP). 
\referenceP is used to extract the quality indicators, as described in \secref{sec:approach}. 
We mitigate this threat by building the \referenceP from a run without the search budget for each case study.
Therefore, \referenceP should contain all the non-dominated solutions across all configurations, and it should also represent a good Pareto front for computing the HV indicator.

\paragraph*{External validity}
Our approach might be affected by \emph{external validity} threats, because we have used a single modeling notation. 
We cannot generalize our results to other modeling notations, which could imply using a different portfolio of refactoring actions. 
In fact, the syntax and semantics of the modeling notation determine the amount and nature of refactoring actions that can be performed. 
We could mitigate these threats, for example, by using another modeling notation.

\paragraph*{Internal Validity}
Our results might be affected by \emph{Internal validity} threats.
One aspect that might affect our findings is a misleading interpretation of the outcome due to the random nature of genetic algorithms.
In order to mitigate the internal validity threats, we performed \independentRun executions for each configuration~\cite{Zitzler_Deb_Thiele_2000}.

\paragraph*{Conclusion validity}
Our results might be affected by \emph{Conclusion validity} threats, since our considerations might change with better-tuned parameters for each algorithm.
We did not perform an extensive tuning phase for each algorithm.
However, we used common parameters to setup the algorithms, which should mitigate these threats~\cite{DBLP:journals/ese/ArcuriF13}.
Wherever possible, we used appropriate statistical procedures with p-value and effect size measures to test the significance of the differences and their magnitude.
  
 \section{Lesson learned}\label{sec:lesson}

We learned that it is not always possible to select a genetic algorithm beforehand.
The selection of the algorithm is strictly related to the domain and the policy for searching the solution space.
Furthermore, the selection becomes even more complex when a search budget limits the search.

Our initial investigation shed light on which algorithm is faster among the studied ones and which algorithm has been able to produce better quality Pareto fronts when the search has a time limitation.
Based on our investigation, we understand that the \nsga policy of non-dominated knowledge helps the algorithm to perform more evolutions. 
In contrast, the hyper-grid searching policy exploited by \pesa allows the algorithm to generate the Pareto fronts with the highest quality.
Finally, \spea, in our experimentation, showed speed and quality limitations due to the usage of two archives for storing Pareto solutions.

We also learned that the application domain is the most driving aspect of genetic algorithms.
In fact, from our results, in one case, we had horizontal clusters (see \ccm on \figref{fig:byalgo_scatter}), which were likely due to the more complex nature of the subject model.
Our results showed a more precise optimization direction for two objectives in the other case.

As a final takeaway, small time budgets could be used in preliminary experiments designed to compare the algorithms and to select the best one for the longer runs.
 \section{Conclusion and Future Work}\label{sec:conclusion}

In this study we presented an investigation on the impact of the search budget for model-based multi-objective refactoring optimization. The study was aimed at helping designers to select the best algorithm with respect to the search budget. 
In addition, we validated the study on two model benchmarks, \texttt{Train Ticket Booking Service}, and \ccm, and on three genetic algorithms, \nsga, \spea, and \pesa.

We assessed the overall quality of each algorithm through the Hypervolume indicator, which measures the amount of the search space volume that a computed Pareto front covers with respect to the reference Pareto front.
From our results, it emerges that \nsga is the fastest algorithm because it generated the highest number of evolution genetic within the search budget. 
\pesa is the algorithm that generated the highest quality results in terms of HV.
Finally, \spea is the slowest algorithm, and it generated the worst quality results. Thus, it generated the lowest number of genetic evolutions, and showed the lowest HV values.

As future work, we intend to analyze the Pareto front at each evolution in order to discover if the quality is not improving enough, and we could just stop the algorithm.
 
\section*{Acknowledgements}
\noindent Daniele Di Pompeo is supported by the Centre of EXcellence on Connected, Geo-Localized and Cybersecure Vehicle (EX-Emerge), funded by the Italian Government under CIPE resolution n. 70/2017 (Aug. 7, 2017).

\noindent Michele Tucci is supported by the OP RDE project No. CZ.02.2.69/\-0.0/\-0.0/\-18\_053/\-0016976 ``International mobility of research, technical and administrative staff at Charles University''.

\bibliographystyle{IEEEtran}
\bibliography{biblio}

\end{document}